# Scaling of an anomalous metal/insulator transition in a 2D system in silicon at $B = 0$

S. V. Kravchenko, Whitney E. Mason, G. E. Bowker*, and J. E. Furneaux
*Laboratory for Electronic Properties of Materials and Department of Physics and Astronomy, University of Oklahoma, Norman, OK 73019*

V. M. Pudalov
*Institute for High Pressure Physics, Troitsk, 142092 Moscow District, Russia*

M. D'Iorio
*National Research Council of Canada, IMS, Ottawa, Ontario, Canada K1A OR6*

We have studied the temperature dependence of resistivity, $\rho$, for a two-dimensional electron system in silicon at low electron densities, $n_s \sim 10^{11}$ cm$^{-2}$, near the metal/insulator transition. The resistivity was empirically found to scale with a single parameter, $T_0$, which approaches zero at some critical electron density, $n_c$, and increases as a power $T_0 \propto |n_s - n_c|^\beta$ with $\beta = 1.6 \pm 0.1$ both in metallic ($n_s > n_c$) and insulating ($n_s < n_c$) regions. This dependence was found to be sample-independent. We have also studied the diagonal resistivity at Landau level filling factor $\nu = 3/2$ where the system is known to be in a metallic state at high magnetic field and in an insulating state at low magnetic field. The temperature dependencies of resistivity at $B = 0$ and at $\nu = 3/2$ were found to be identical. These behaviors suggest a true metal/insulator transition in the two dimensional electron system in silicon at $B = 0$, in contrast with the well-known scaling theory.

## I. INTRODUCTION

For a number of years it has been generally believed that at zero magnetic field, all the states are localized in the two dimensional electron system (2DES) in the limit of infinite sample size. Arguments based on scaling theory indicate that as $T \to 0$, resistivity always increases, exponentially in the case of "strong" localization or logarithmically in the case of "weak" localization [1]. Early experiments on relatively low mobility samples [2,3] confirmed this behavior. Thus, conventional wisdom has been that there is no metal/insulator (M/I) phase transition in an infinite 2D sample because there can be no metal, in contrast with the three dimensional situation where electrons are localized only if the Fermi energy, $E_F$, lies below some mobility edge, $E_c$. Recently, however, there have been a number of experimental and theoretical results which have lead us to study this problem further.

Experiments at high magnetic fields associated with the quantum Hall effect, QHE, clearly show (see, *e.g.*, Ref. [4]) extended states at $E_F$ when the Landau filling factor $\nu = n_s hc/eB = i + 1/2$ (here $n_s$ is the 2D electron density, $h$ is the Plank constant, $c$ is the speed of light, $e$ is the electron charge, $B$ is magnetic field, and $i$ is an integer). Recently there has been considerable interest in the transition from QHE behavior to insulator behavior at lower magnetic fields [5–8]. Here, theory indicates that the extended states will float up in energy as $B \to 0$ leading to an insulating state [9,10]. Experimentally, the case is not so clear; Shashkin *et al.* [7] have found recently that the extended states coalesce and float up as $B \to 0$ but do not necessarily lead to an insulator. New theoretical arguments [11] have also fueled a re-examination of the behavior of a 2DES at $B = 0$. In Ref. [11], Azbel finds that for a system of noninteracting 2D electrons in a model disorder potential with a random set of special scatterers at $B = 0$ is localized *only* when $E_F < E_c$. At all energies above $E_c$, extended states exist. According to Azbel, the disagreement between his results and those of Abrahams and coworkers [1] might indicate that the resistance strongly depends on the range of the scattering centers. In the dilute 2D electron system in silicon metal-oxide-semiconductor field-effect transistors (MOSFET's), scatterers are short-range, similar to the model potential used by Azbel, and thus his results may be applicable.

For these reasons we recently studied [12] high-mobility Si MOSFET samples, and showed that conventional weak localization, observed at $T \gtrsim 1 - 2$ K, is overpowered by an order of magnitude drop in resistivity, $\rho$, as the temperature is decreased below $\sim 1$ K. No signs of electron localization are seen down to the lowest available temperature (20 mK) even for very low electron densities above some critical value, $n_c$. At electron densities lower than this critical value, the resistivity monotonically increases as $T \to 0$, indicating a localized state studied extensively elsewhere [13]. Moreover, the resistance is empirically found to scale with temperature at densities both below and above $n_c$ [12]. Mathematically this indicates that the resistivity can be written in the form

$$\rho(T, n_s) = \rho[T/T_0(n_s)]. \tag{1}$$

The scaling parameter, $T_0$, has been found to approach zero at $n_s = n_c$. The observed behaviors resemble an M/I transition in three dimensions rather than the behavior expected for a 2D system [14], and suggest a true M/I transition in Si MOSFET's at zero magnetic field, in contrast with predictions of Ref. [1] and consistent with Ref. [11]. It should be mentioned that an analogous temperature behavior of resistivity and scaling on both sides of the transition point was observed in thin disordered Bi films where it was considered as evidence for a direct phase transition from an insulating state to a superconducting state [15].

Here we report extensive studies of scaling behavior in high-mobility 2DES in silicon at $B = 0$. We show that the dependence of the scaling parameter, $T_0$, on electron density is symmetric about $n_c$ and obeys a power law

$$T_0 = C|\delta_n|^\beta \text{ with } \beta = 1.6 \pm 0.1 \tag{2}$$

where $\delta_n = n_s - n_c$, both for insulating and metallic sides of the transition. Both the power $\beta$ and the coefficient $C$ were found to be essentially sample-independent. We related $T_0$ for the insulating side to a localization length $\xi$ using the equation $k_B T_0 = e^2/\epsilon\xi$ (where $\epsilon$ is the dielectric constant) because we found $\rho(T) \propto \exp\,(T_0/T)^{1/2}$ which is characteristic for a Coulomb gap [17]. The dependence of $\rho$ on the electron density close to the transition point for different temperatures appeared to be qualitatively the same as those for QHE/insulator transition [6,8], with one temperature independent crossing point (see Fig. 1). We have also performed temperature studies of the diagonal magnetoresistivity, $\rho_{xx}$, at $\nu = 3/2$ where a 2DES is known [4] to be in a true metallic state in the high $n_s$/high $B$ limit and in an insulating state in the low $n_s$/low $B$ limit. Therefore, at $\nu = 3/2$, there must exist a true M/I transition observable at constant $\nu$ with decreasing $B$ and $n_s$ as shown schematically for the phase diagram of Refs. [5,7,16] in Fig. 2. We have found the temperature and density behavior of $\rho$ at $B = 0$ to be absolutely identical to that for $\rho_{xx}$ at $\nu = 3/2$, providing additional evidence for a true metal/insulator phase transition in zero magnetic field.

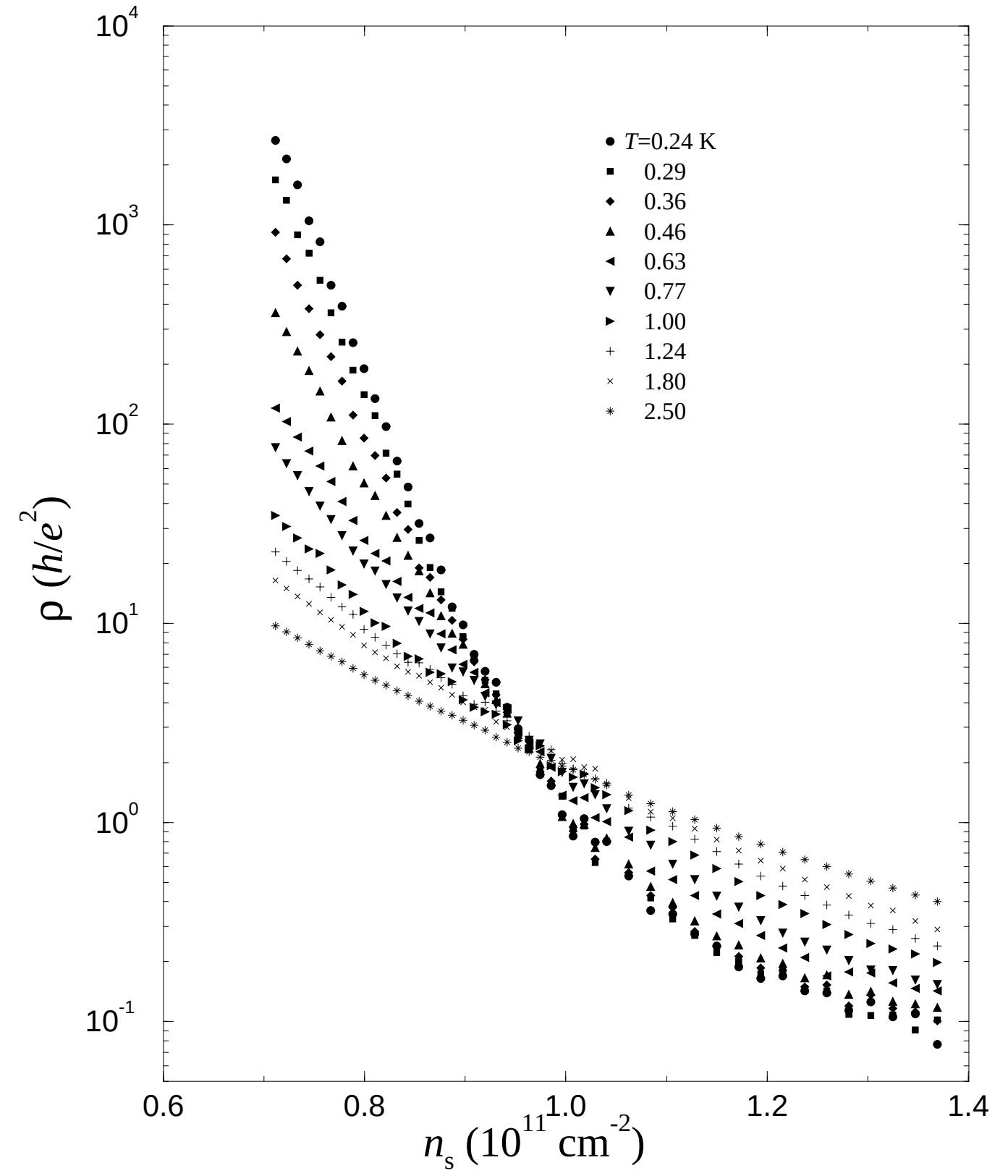

FIG. 1. Dependencies of $\rho$ in units of $h/e^2$ on electron density for different temperatures at $B = 0$. Sample Si-12b.

## II. EXPERIMENTAL

Below we show results obtained with three samples from two wafers: Si-12a with maximum mobility, $\mu_{max}$, of $3.5 \times 10^4$ cm$^2$/Vs, Si-12b with $\mu_{max} = 3.0 \times 10^4$ cm$^2$/Vs, and Si-15 with $\mu_{max} = 7.1 \times 10^4$ cm$^2$/Vs. Other samples from other wafers showed similar results. The samples are rectangular Hall bars with a source to drain length of 5 mm, a width of 0.8 mm, and an intercontact distance of 1.25 mm. The thickness of the oxide separating 2DES and the gate was close to 2000 $\AA$ for all samples. The resistance was measured using a four-terminal dc technique with cold amplifiers (input resistances $> 10^{14}$ Ω) installed on the 1 K pot of a dilution refrigerator. The output of these amplifiers was connected to a standard digital voltmeter. Great care was taken to ensure that all data were obtained in the region of linear $I - V$ characteristics. To accomplish this, it was necessary to chose a proper measuring current which varied from several pA for high resistances to 100 nA for low resistances. This was especially important at low temperatures where $\rho$ varies by 5 orders of magnitude within the chosen $n_s$ interval. For each sample we observed the same $\rho(T, n_s)$ characteristics independent of contact configuration. Samples were mounted with a weak thermal link to the mixing chamber (via a stainless steel rod) allowing a change in the temperature from 0.2 to 7.5 K during the experiment. We controlled the temperature using two calibrated resistance thermometers placed in good thermal contact with the sample.

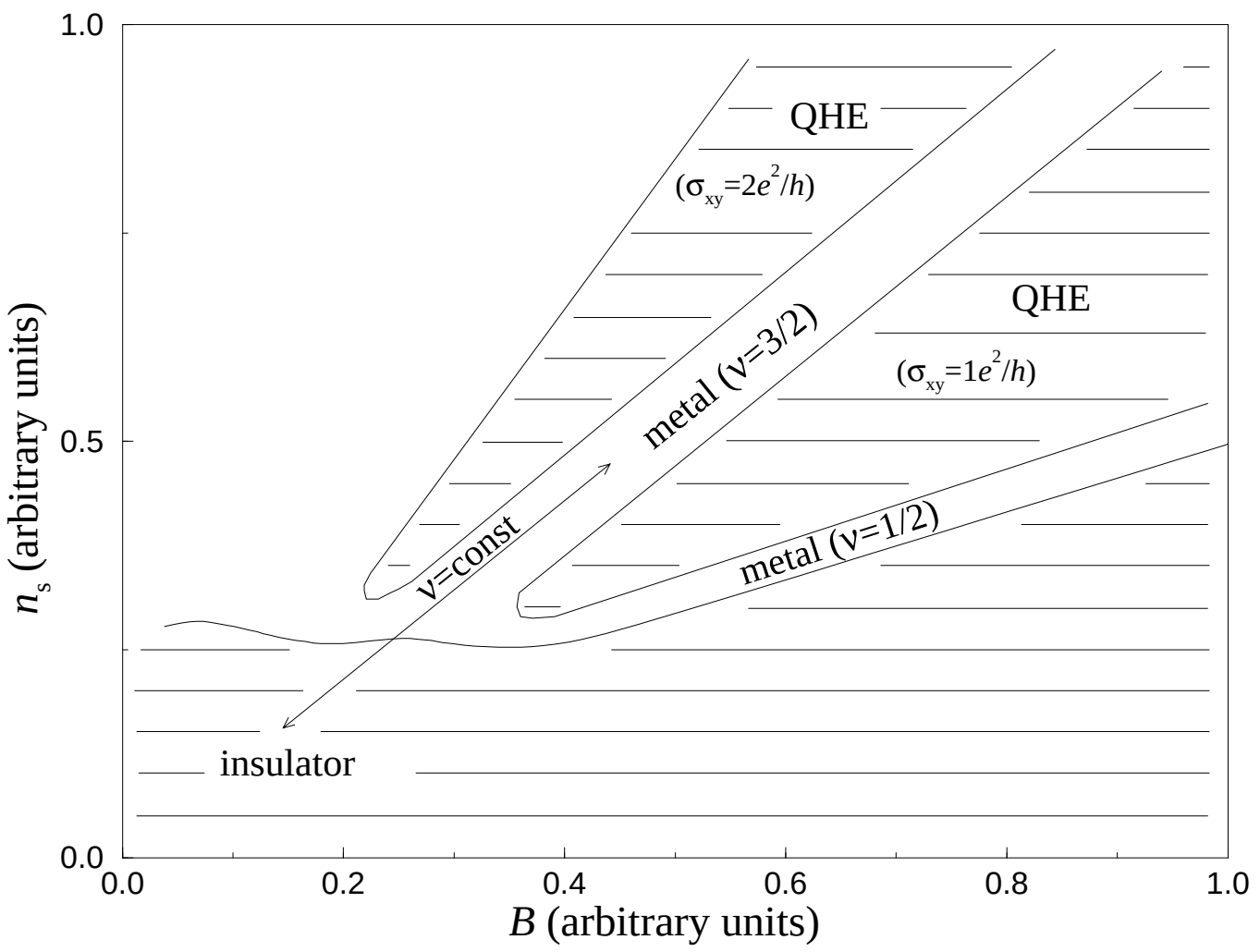

FIG. 2. Schematically shown metal/insulator phase diagram in the $(B, n_s)$ plane taken from Refs. [5,7,16]. Lower shaded region corresponds to the insulating state, two upper shaded regions — to the quantum Hall effect with $\sigma_{xy} = 1\,e^2/h$ and $2\,e^2/h$. Blank strips between shaded regions correspond to metallic states. The arrow shows the metal/insulator transition at $\nu = 3/2$ realized in the current experiment.

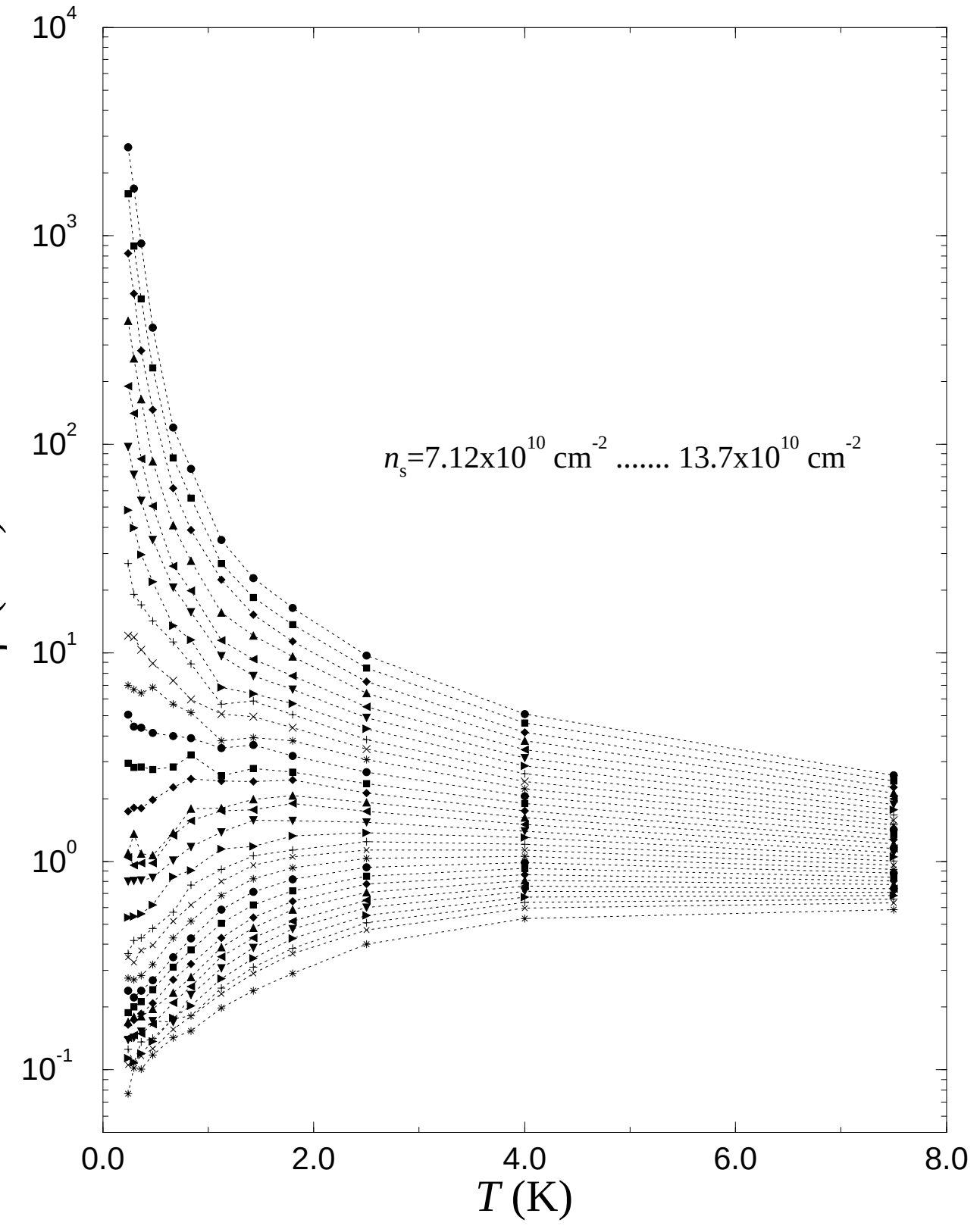

FIG. 3. Temperature dependencies of the resistivity (sample Si-12b) for different electron densities at $B = 0$.

Figure 1 shows the resistivity (in units of $h/e^2$) as a function of electron density for Si-12b for several temperatures. One can see that all curves cross at some resistivity, $\rho^0 \sim 2h/e^2$, and electron density, $n_c = 0.96 \times 10^{11}$ cm$^{-2}$, which corresponds to a mobility of about $10^3$ cm$^2$/Vs. At densities below this point, the resistivity is higher for lower temperatures, behavior which is characteristic of an insulating state. In contrast, for $n_s > n_c$, the lower the temperature, the lower the resistance, behavior which is characteristic of a metallic state. The data, including $\rho^0$, for other samples are identical except $n_c$ varies (see also Ref. [12]). This behavior, particularly the existence of a single crossing point, is qualitatively identical to the behavior of the QHE to insulator transition (see Figs. 2 and 4 in Ref. [6] and Fig. 3 in Ref. [8]).

To see the temperature dependence of resistivity, in Fig. 3 we replot $\rho$ data as a function of temperature for 30 different electron densities varying from $7.12 \times 10^{10}$ to $13.7 \times 10^{10}$ cm$^{-2}$. At low densities, the curves grow monotonically as the temperature decreases, behavior characteristic of an insulator [18]. However, for $n_s \gtrsim n_c$, the temperature behavior of $\rho$ becomes nonmonotonic: resistivity increases at $T \gtrsim 2$ K and decreases as the temperature is decreased; this behavior is "insulating" at higher $T$ and "metallic" at lower $T$. At still higher $n_s$, resistivity is almost constant at $T \gtrsim 4$ K but falls by an order of magnitude at lower temperatures showing a strongly metallic behavior as $T \to 0$.

A striking feature of the $\rho(T)$ dependencies for different $n_s$ is that they can be made to overlap by scaling them along the $T$-axis. In other words, resistivity can be represented as a function of $T/T_0$ with $T_0$ depending only on $n_s$. This was possible for quite a wide range of electron densities (typically $n_c - 2.5 \times 10^{10} \lesssim n_s \lesssim n_c + 2.5 \times 10^{10}$ cm$^{-2}$) and in the temperature interval 0.2 to 3 K. The results of this scaling are shown in Fig. 4 where $\rho$ is represented as a function of $T/T_0$. One can see that the data dramatically collapse into two separate curves, the upper one with open symbols for the insulating side of the transition and the lower one with filled symbols for the metallic side. The thickness of the lines is largely governed by the noise within a given data set which indicates the high quality of the scaling. It is worth noting that qualitatively the same scaling picture was obtained in Ref. [15] for insulator/superconductor transition in thin Bi films.

The procedure used to bring about the collapse and determine $T_0$ for each $n_s$ was the following. First, we noted (see Fig. 5) that at low $n_s$, to high precision the resistivity has a temperature behavior characteristic of hopping in the presence of a Coulomb gap [17]:

$$\rho = \rho_0 \exp\left[(T_0/T)^{1/2}\right]. \quad (3)$$

Therefore, $T_0^{(1)}$ for the lowest $n_s$ (*i.e.*, for the upper curve in Fig. 3) was determined by fitting Eq. 3 to the resistance data. Next, the second curve from the top in Fig. 3 was scaled along the $T$-axis with the factor $\gamma^{(2)}$ to coincide with the upper curve; $T_0$ for the second curve thus was determined as $T_0^{(2)} = \gamma^{(2)} T_0^{(1)}$. Then the third curve from the top was scaled with the factor $\gamma^{(3)}$ to coincide with the two upper ones determining $T_0^{(3)} = \gamma^{(3)} T_0^{(1)}$. This procedure was repeated subsequently for all "insulating" curves ($n_s < n_c$, $d\rho/dT < 0$) in Fig. 3. This gives the upper curve in Fig. 4 (a) designated with open symbols. The scaling parameter, $T_0$, for the insulating side of the transition, is shown in Fig. 4 (b) by open symbols as a function of electron density. One can see that it approaches zero as $n_s$ approaches the critical point, $n_c$. Identical behavior for a second sample is shown in Fig. 6.

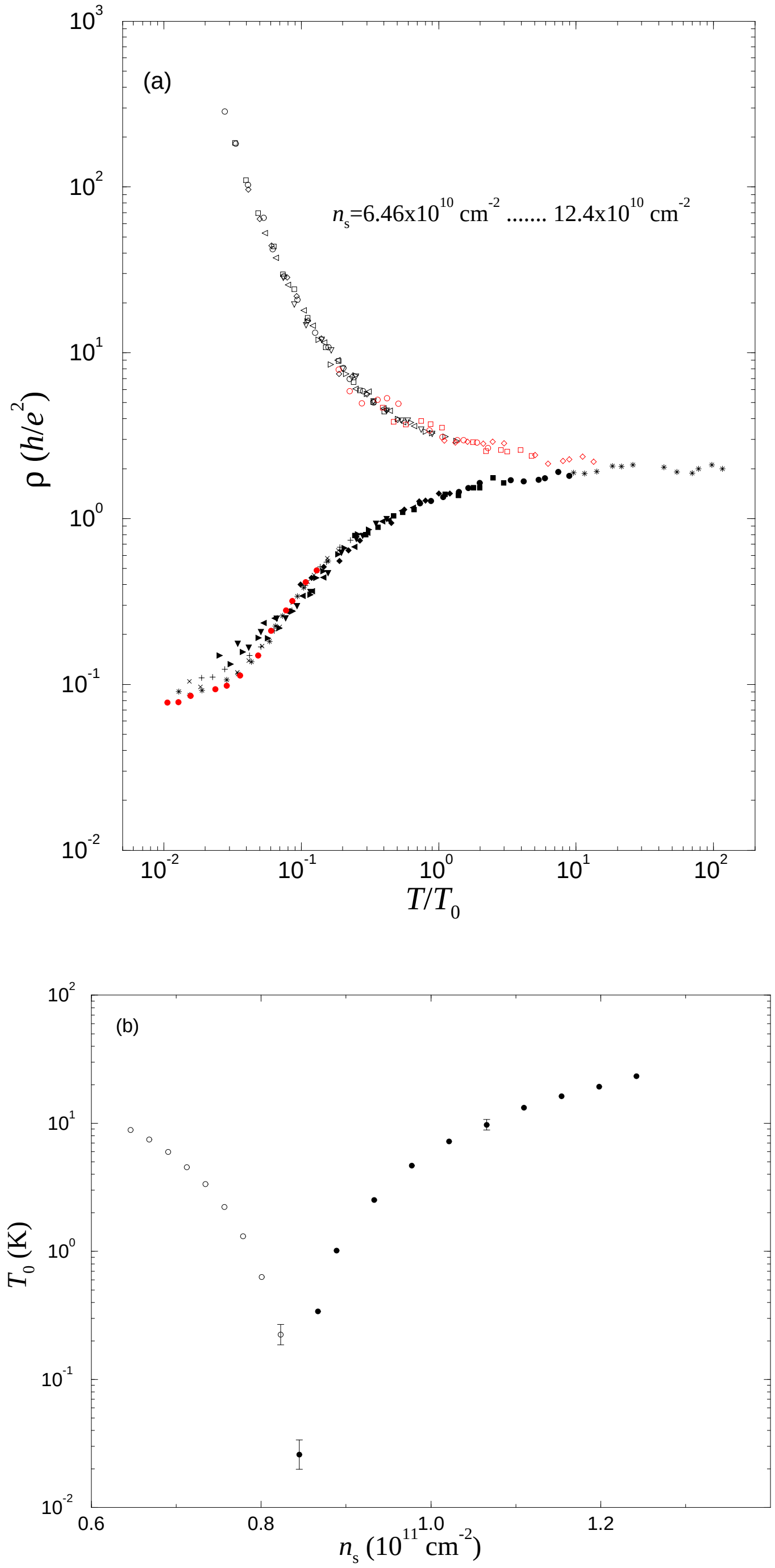

FIG. 4. Resistivity vs $T/T_0$ (a) and scaling parameter, $T_0$, vs electron density (b) for Si-12a. Open symbols correspond to the insulating side of the transition, closed — to the metallic one.

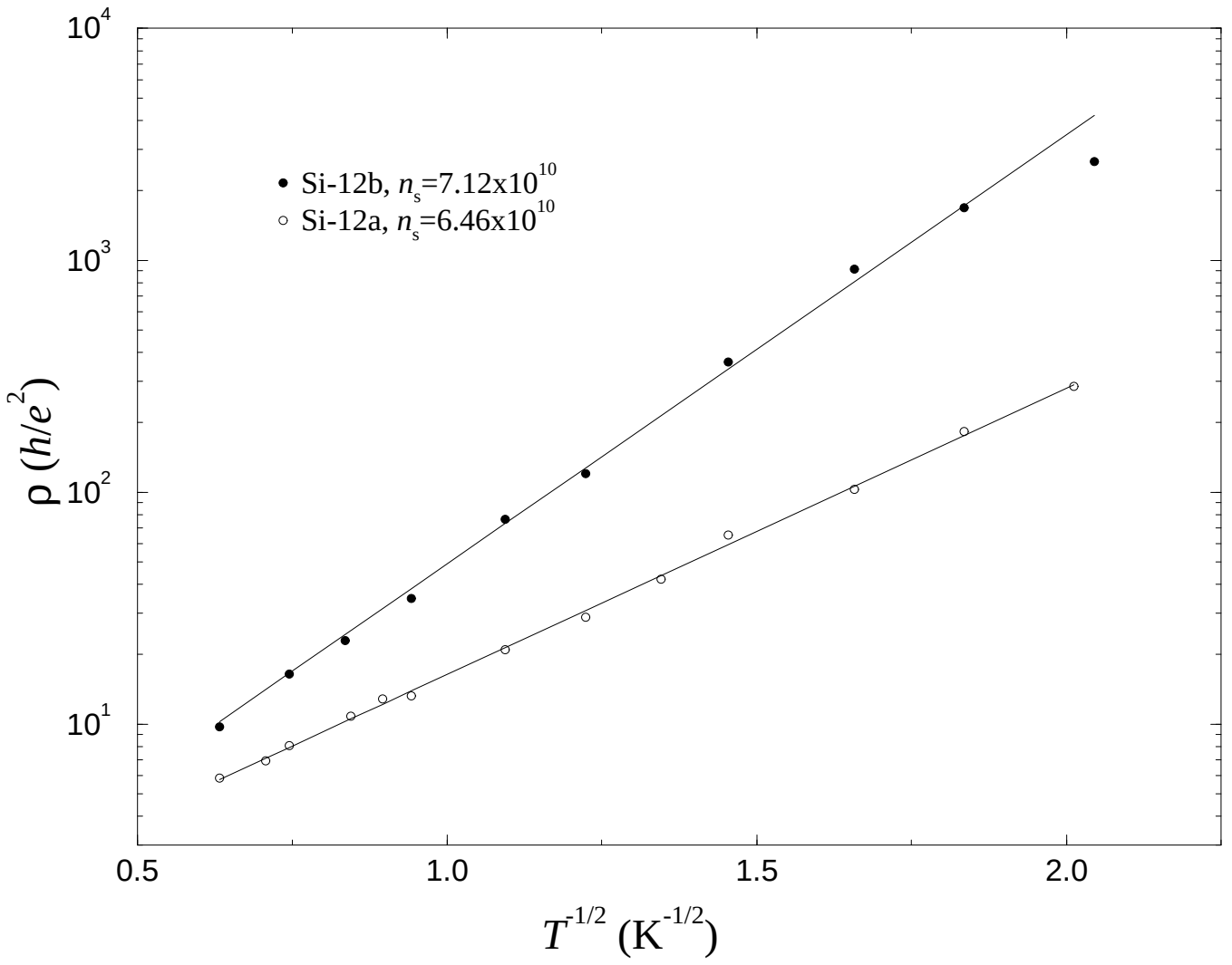

FIG. 5. Resistivity vs $T^{-1/2}$ for two samples at lowest electron densities which were achieved.

To collapse the "metallic" curves ($n_s > n_c$) into a single curve, we applied the same procedure. We started with the curve corresponding to the highest $n_s$ (*i.e.*, with the lowest curve in Fig. 3) and then scaled the $T$-axis for the other curves by factors $\gamma^{(2)}$, $\gamma^{(3)}$ *etc* to make them coincide with the first curve. To fix the position of the resulting curve in the $\log(T/T_0)$ axis, it is necessary to assign some value of $T_0$ to the first curve, and then get $T_0$ for other curves as $T_0^{(k)} = \gamma^{(k)} T_0^{(1)}$. In contrast to the "insulating side", it is not clear what value of $T_0$ should be assigned to the first metallic curve. We have found (see Fig. 7) that for both the insulating side and the metallic side of the transition, the dependence of $\gamma$ on $|\delta_n| = |n_s - n_c|$ is a power law:

$$\gamma(\delta_n) \propto |\delta_n|^\beta, \tag{4}$$

with the same value of $\beta$ for metallic and insulating sides. This common power law can be clearly seen in Fig. 7 where for each sample the open (insulating side) and filled (metallic side) symbols form a single line. Because $\gamma(|\delta_n|)$ has the same behavior on both sides of the transition, we suggest that the function $T_0(|\delta_n|)$ is also quantitatively symmetric. In Fig. 10 and in the discussion below we further justify this assumption [19]. The scaling factor as a function of electron density for the metallic side of the transition is shown in Figs. 4 (b) and 6 (b) by closed symbols. Again, as for the insulating side, $T_0 \to 0$ as $n_s \to n_c$.

Figure 7 shows $T_0$ (in log-log format) for both metallic and insulating sides of the transition for three different samples as a function $|\delta_n|$. As we mentioned above, the dependency $T_0(|\delta_n|)$ is a power law. We find the average power $\beta$ to be $1.60 \pm 0.1$ for insulating side of the transition and $1.62 \pm 0.1$ for metallic side. For all three samples shown in the figure, the dependencies $T_0(|\delta_n|)$ are nearly identical in spite of the fact that the values of $n_c$ and sample mobilities are different. We should mention that the similar power law with an exponent $1.5 \pm 0.2$ was observed for scaling of the superconductor/insulator transition in thin Bi films [15,20] though we believe the physical mechanism driving the transition is different in the two systems.

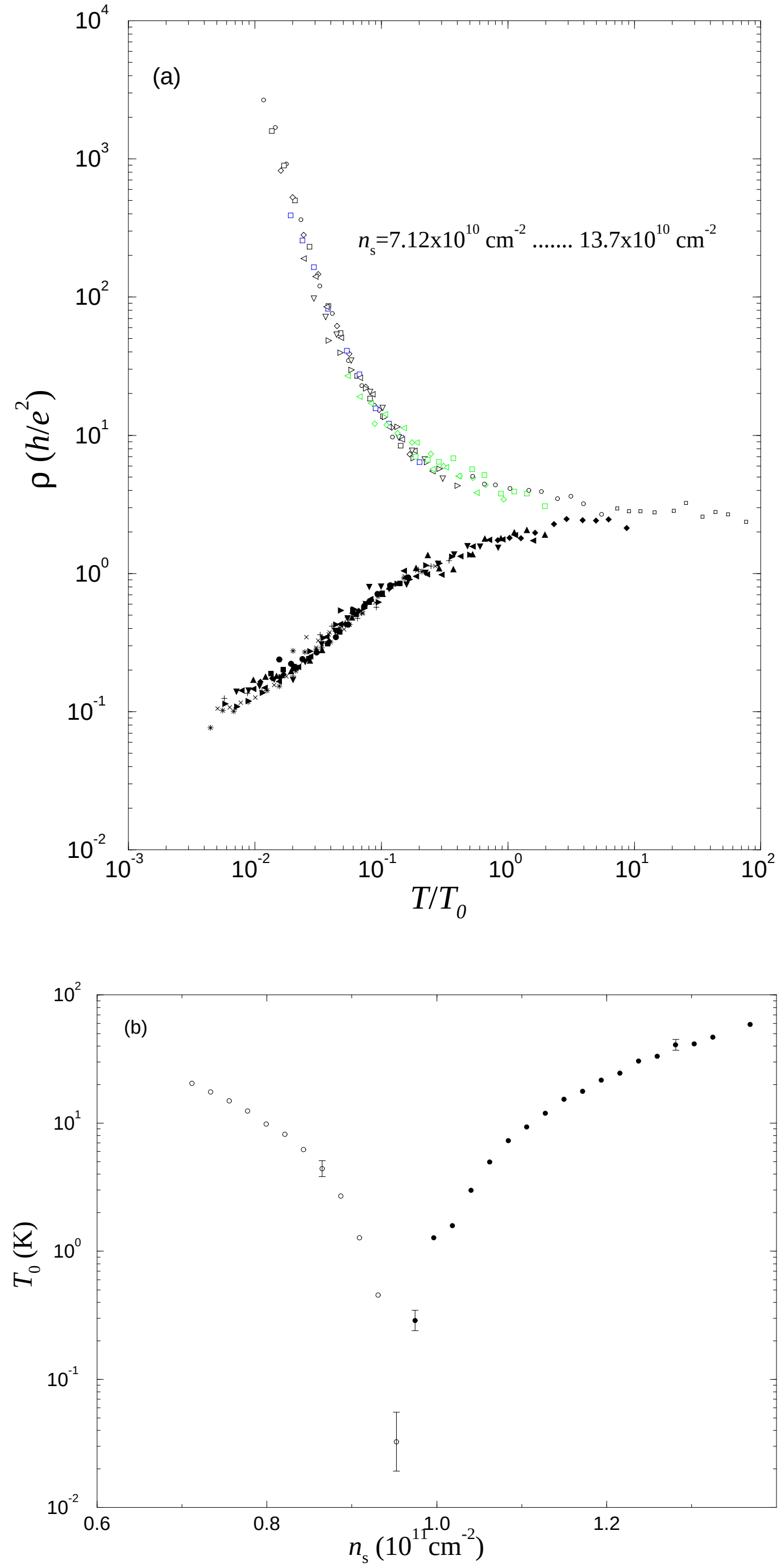

FIG. 6. The same as Fig. 4 but for Si-12b.

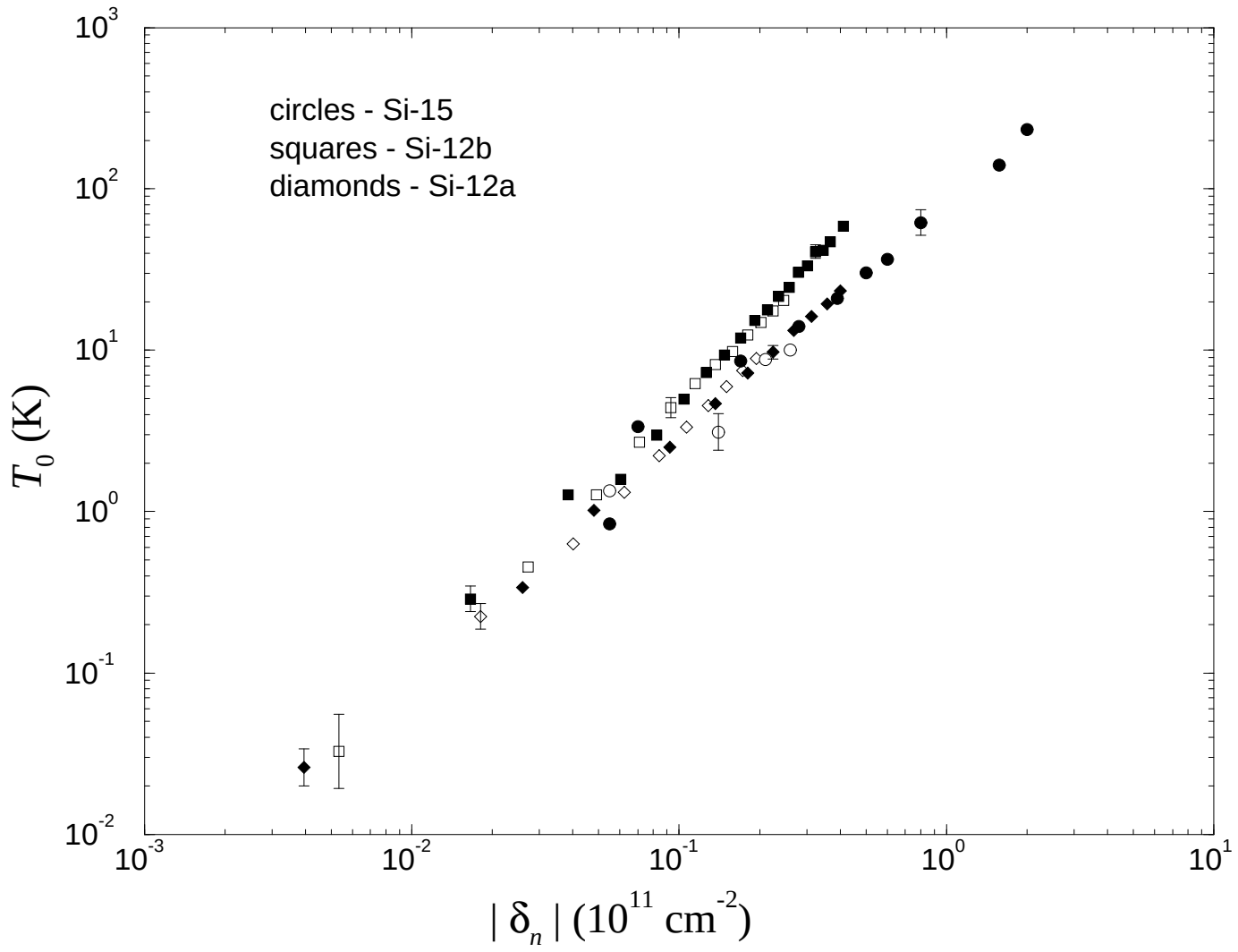

FIG. 7. Scaling parameter, $T_0$, as a function of $|\delta_n|$ for three samples from different wafers.

Finally, we show the temperature dependencies of the diagonal resistivity, $\rho_{xx}$, in a magnetic field corresponding to a Landau level filling factor $\nu = 3/2$. To obtain this data, we varied both $n_s$ and $B$ so that $\nu$ remained constant as shown in Fig. 2. Along this path one expects a true M/I transition. Figure 8 shows $\rho_{xx}(T)$ dependencies for $\nu = 3/2$ (symbols) along with $\rho(T)$ dependencies at $B = 0$ (dashed lines). Surprisingly, the curves are practically identical even though $n_c$ is different. In particular, curves with high-$T$ resistivities of similar values were found to have identical temperature dependencies indicating that $\rho^0$ is also the same. This shows that the M/I transitions at zero magnetic field and at half-integer filling factor are identical when viewed in terms of resistivities.

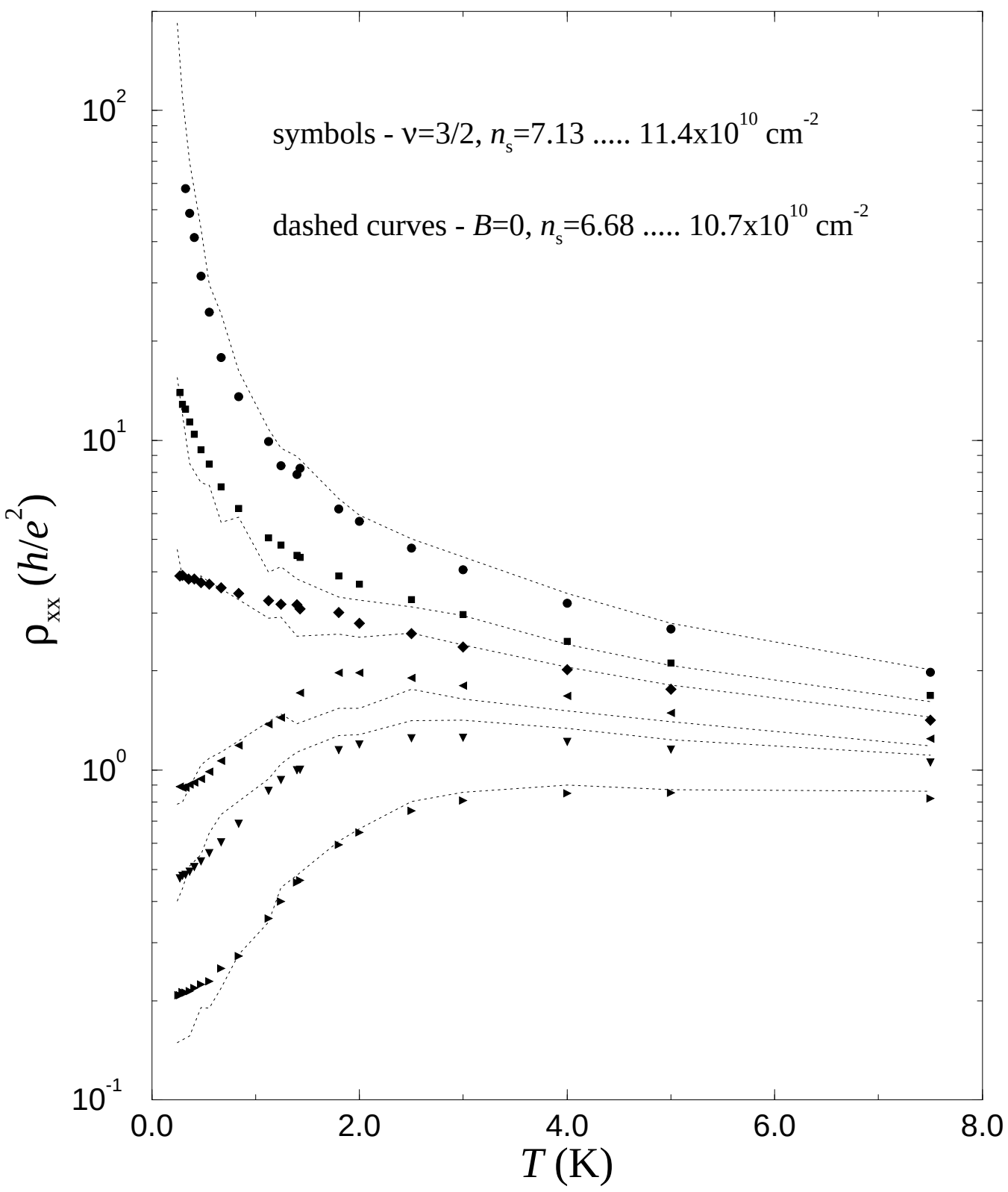

FIG. 8. Resistivity vs temperature for Si-12a in a magnetic field corresponding to $\nu = 3/2$ (symbols) and at $B = 0$ (dashed curves) for different electron densities.

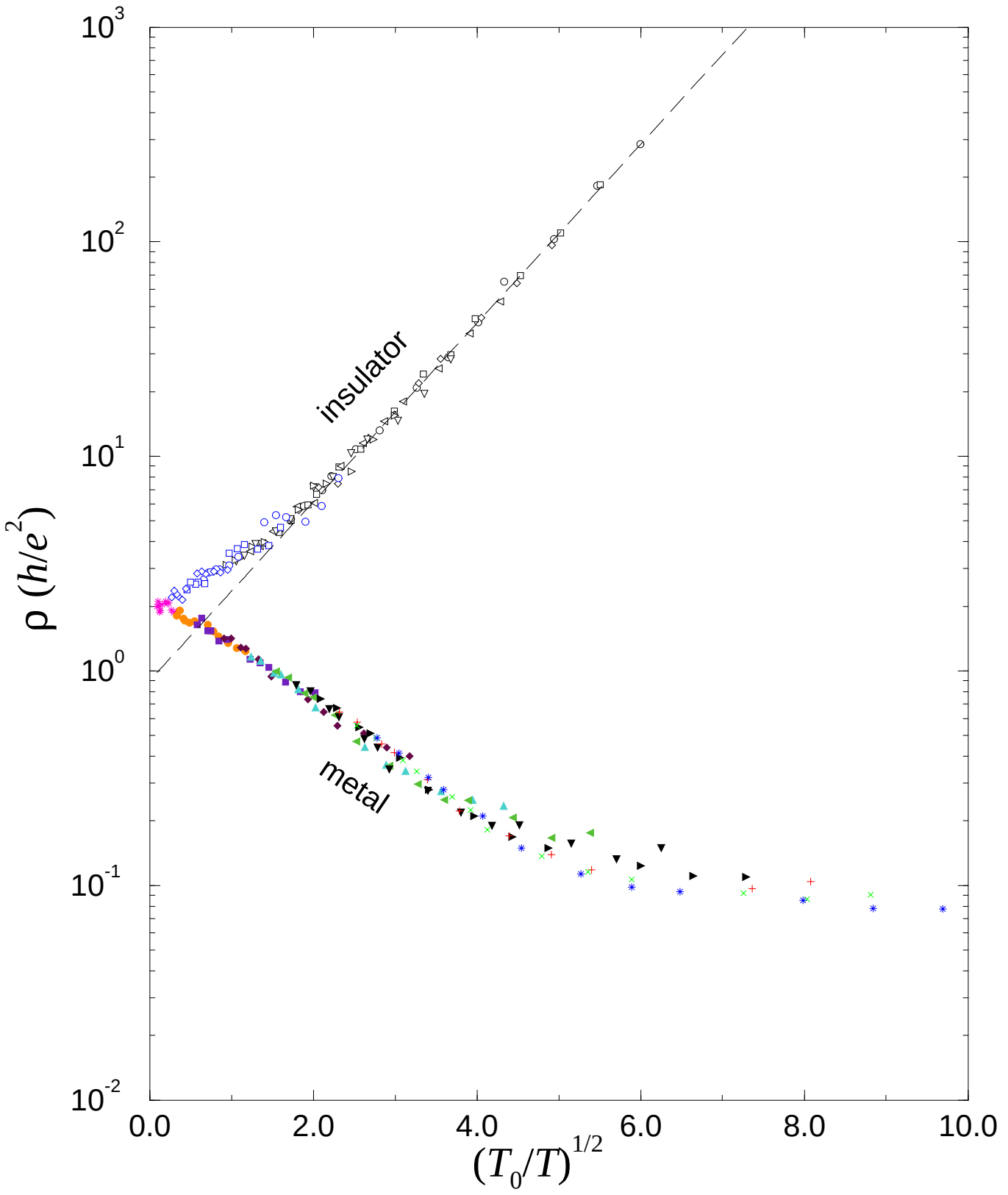

FIG. 9. Resistivity vs $T/T_0$ for Si-12a. Open symbols — the insulating side of the transition; closed — the metallic one.

## III. DISCUSSION

Scaling of an appropriate physical variable is one of the hallmarks of a phase transition. In this paper we report the scaling behavior for the resistivity of the high-quality 2DES in Si MOSFET's with the scaling parameter approaching zero as the density of electrons approaches the critical value. This behavior strongly suggests a true metal/insulator transition in the two-dimensional system in silicon at zero magnetic field, in apparent contradiction to the theoretical arguments of Abrahams and coworkers [1] and consistent with those of Azbel [11]. We must note however that the temperature behavior of resistivity on the metallic side of the transition with $\rho$ dropping by an order of magnitude at temperatures below 1–2 K is not what one would expect for an ordinary metal where resistivity saturates when the frequency of phonons $\Theta = k_BT/h$ becomes less than $\tau^{-1}$, the inverse elastic scattering time, *i.e.*, at temperatures $T \lesssim T_\Theta = h/k_B\tau$ (in the regime of interest, $T_\Theta > 10$ K). We have discussed in detail the precipitous drop of resistivity at $T < 1-2$ K in Ref. [12]. In particular, we showed that this drop can be explained neither by electron-phonon scattering nor by the temperature-dependent screening considered in Refs. [21,22].

In Fig. 9 we plot the resistivity as a function of $[T_0/T(n_s)]^{1/2}$. The resulting "metallic" curve is shown by closed symbols (lower curve). It monotonically decreases by more than an order of magnitude as $[T_0/T(n_s)]^{1/2}$ increases and finally saturates at $[T_0/T(n_s)]^{1/2} \gtrsim 6$. The upper, "insulating" curve (open symbols) at $(T/T_0)^{-1/2} \gtrsim 2$ can be fitted well by the formula

$$\rho = A\,\exp[(T/T_0)^{-1/2}] \tag{5}$$

where the prefactor $A$ is close to $h/e^2$. At $(T_0/T)^{1/2} \lesssim 2$ (*i.e.*, when $T$ becomes close to $T_0$) the dependence weakens. As we have already mentioned, this behavior is characteristic of hopping in the presence of a Coulomb gap [17]. This contradicts earlier experimental results of Ref. [23] where the absence of the Coulomb gap in Si MOSFET's was reported. We attribute the difference to the greater role of Coulomb effects relative to scattering by impurities in our samples because our samples are of higher mobility (and therefore are less disordered) at low electron densities. Recently, Coulomb gap behavior of the resistivity was reported for low-density gated GaAs/AlGaAs heterostructures [24]. In this system, a transition from the Coulomb gap behavior to the Mott variable range hopping behavior [25], $\rho \propto \exp\,[(T/T_0)^{-1/3}]$, was observed as the temperature was lowered, in agreement with recent theoretical predictions [26]. In principle, in our system, $\rho(T)$ characteristic of the Mott variable range hopping regime can develop at lower temperatures and lower electron densities leading to the destruction of scaling as $T \to 0$.

Qualitatively the same $\rho(T/T_0)$ function was observed for the insulating state in thin palladium films [27]. The resistivity for the thinnest films also was found to obey the Coulomb gap law. However, the system studied in Ref. [27] was always weakly or strongly insulating; there was no "metallic part" of the scaling function.

We obtain physical insight into the observed scaling by relating the scaling parameter, $T_0$, to a length scale as indicated in Ref. [17],

$$k_BT_0 = \frac{e^2}{\epsilon\xi}. \tag{6}$$

Therefore, if we assume that $\epsilon$ is constant [28], the approach of $T_0$ to zero as $\delta_n \to 0$

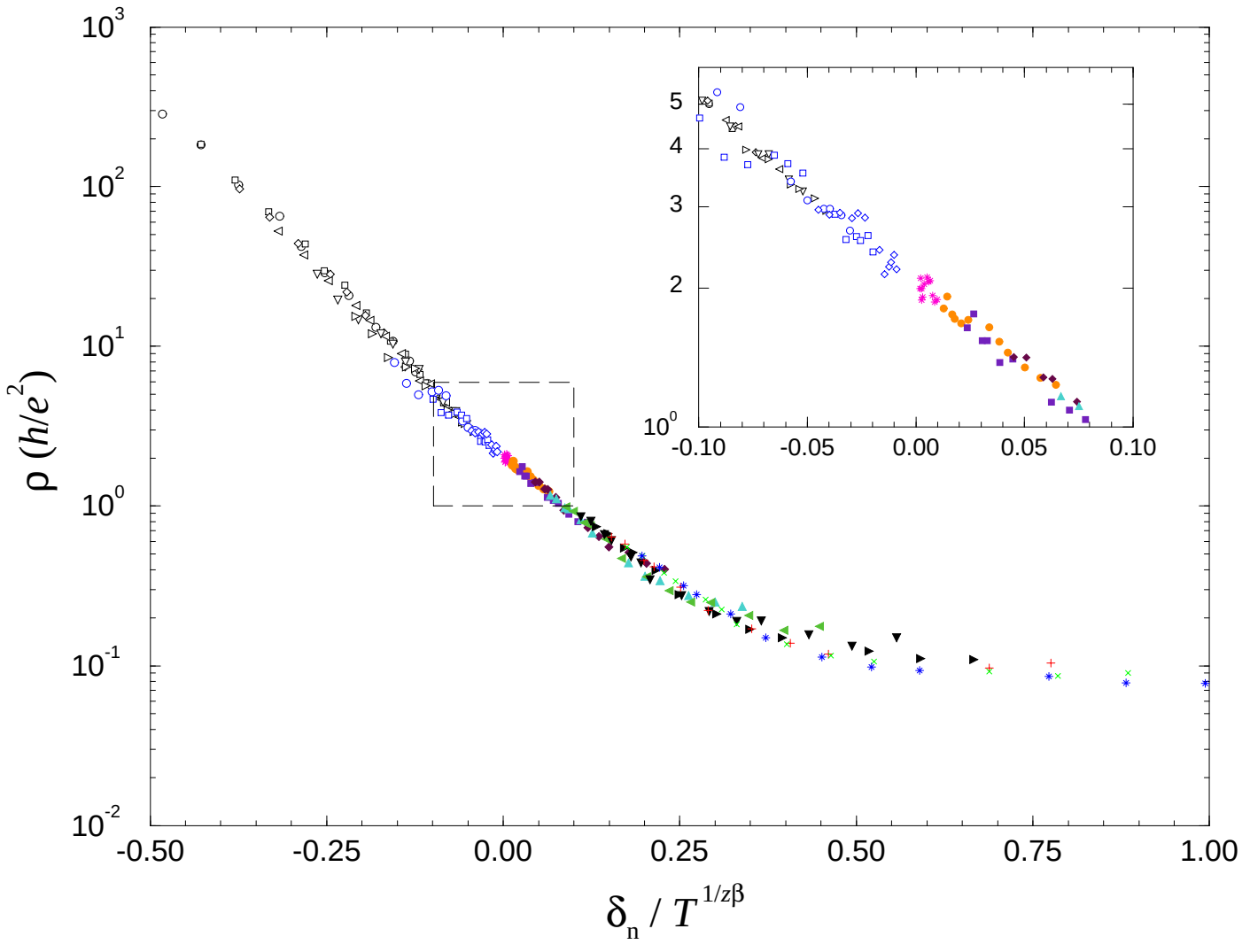

FIG. 10. Resistivity vs the scaling variable, $\delta_n/T^{1/z\beta}$ with $z = 1$ and $\beta = 1.6$. Open symbols — the insulating side of the transition; closed — the metallic one.

indicates a diverging localization length:

$$\xi \propto |\delta_n|^{-\beta} \text{ with } \beta \approx 1.6 \tag{7}$$

(the conventional symbol for this exponent is $\nu$, but we have used $\beta$ in this paper to avoid confusion with $\nu$, the Landau filling factor). Similar behavior for the localization length (with $\beta \approx 1$) was recently deduced from nonlinear current-voltage characteristics [16].

Figure 10 shows $\rho$ as a function of the scaling variable, $X = \delta_n/T^{1/z\beta}$, used in Ref. [29] for the description of the superconductor/insulator transition. Here $z$ is the dynamical exponent equal to 1 in the case of purely Coulombic interaction with constant $\epsilon$. The exponent $\beta$ was taken to be 1.6. Again, all the data collapse onto a single curve. One can see that $\rho$ as a function of the scaling variable, $X$, is continuous and smooth around the transition point, $X = 0$ (see inset). This indicates that the mechanism responsible for the temperature dependence of resistivity on both insulating and metallic sides of the transition is the same, and justifies our assumption about qualitative symmetry of the scaling parameter, $T_0$, around the critical point.

Let us now discuss $\rho_{xx}(T)$ for filling factor $\nu = 3/2$. In contrast with the zero magnetic field situation, there is no doubt that in high enough magnetic fields there are extended states at the center of each Landau level [4] corresponding to Landau level filling factors $\nu = i + 1/2$. At low magnetic fields, the extended states no longer follow the centers of the Landau levels but are expected to float up in energy [9,10]. It was recently shown experimentally [7] that the extended states indeed float up in energy at low magnetic fields. In contrast with theoretical expectation, however, their energy does not increase infinitely; instead they combine at some finite energy. An insulator/metal/QHE phase diagram for high-mobility Si MOSFET's (similar to those used in this work) was obtained in Refs. [5,7,16], and is shown schematically in Fig. 2. One can see metallic strips at $\nu = 1/2$ and $3/2$, corresponding to the extended states at these filling factors, between quantum Hall states with $\sigma_{xy} = 1\,e^2/h$ and $2\,e^2/h$ and insulating state with $\sigma_{xy} = 0$ (metallic strips at other half-integer filling factors are not shown). Staying at the same filling factor $\nu = 3/2$, therefore, one can observe a transition from metallic state in the high $B$/high $n_s$ limit to the insulating state at lower $B$ and $n_s$.

Temperature dependencies of the diagonal resistivities shown in Fig. 8 for both $\nu = 3/2$ (symbols) and zero magnetic field (dashed lines) are essentially identical. The curves having close "starting resistivities" at high temperatures have the same temperature dependencies. This strongly suggests that the M/I transition in zero magnetic field is identical to the M/I transition at half-integer filling factor. We note that the identical M/I transitions at $B = 0$ and at high $B$ were indicated in Ref. [16] where nearly field-independent behavior of the localization length was reported.

## IV. SUMMARY

In summary, we have shown that the zero magnetic field resistivity of the high-quality two-dimensional electron system in silicon scales with temperature. A single scaling parameter collapses the resistivity data onto two curves, insulating for $n_s < n_c$ and metallic for $n_s > n_c$, and decreases upon approaching the critical electron density as a power $\beta = 1.6 \pm 0.1$ for both metallic and insulating sides of the transition. This scaling behavior strongly suggests a metal/insulator phase transition for the 2DES in Si MOSFET's at $B = 0$. We have compared this transition with the phase transition in a magnetic field corresponding to a constant Landau level filling factor $\nu = 3/2$ and with the superconductor/insulator phase transition in thin Bi films. The behavior we observe is identical to the $\nu = 3/2$ case and quite similar to the Bi case. These similarities add further weight to our identification of the reported transition as a true M/I phase transition. In particular, the identical behavior at $\nu = 3/2$ and at $B = 0$ along with the arguments presented in Ref. [12] indicate that the observed scaling behavior is not due to some sample size effect associated with weak localization. Thus, our data appears to be consistent with the work of Azbel [11], while the conclusions of Ref. [1] seem inappropriate for high-mobility Si MOSFET's. We are in the process of extending our measurements to GaAs/AlGaAs samples, of studying the scaling behavior at low $B$ and with electric fields.

## ACKNOWLEDGMENTS

We acknowledge useful discussions with B. Shklovskii, A. M. Goldman, L. Zhang, B. Mason, K. Mullen, S. Simon, X. C. Xie and experimental assistance from K. Schneider. This work was supported by grants DMR 89-22222 and Oklahoma EPSCoR via LEPM from the National Science Foundation, grant 93-0214235 from the Russian Fundamental Science Foundation, a Natural Sciences and Engineering Research Council of Canada (NSERC) operating grant, and a grant from the Netherlands Organization for Science, NWO.

---

* Permanent address: Department of Physics and Astronomy, Bowdoin College, Brunswick, ME 04011.